
\magnification=1200
\parskip=\medskipamount \overfullrule=0pt
\parindent=20truept
\font \titlefont=cmr10 scaled \magstep3
\font \namefont=cmr10 scaled \magstep1

\def\singlespace{\baselineskip=\normalbaselineskip}

\newcount\firstpageno \firstpageno=2
\footline={\ifnum\pageno<\firstpageno{\hfil}\else{\hfil
                                                  \rm\folio\hfil}\fi}
\def\frac#1/#2{\leavevmode\kern.1em
 \raise.5ex\hbox{\the\scriptfont0 #1}\kern-.1em
 /\kern-.15em\lower.25ex\hbox{\the\scriptfont0 #2}}

\def\pp{\par\hangindent=.125truein \hangafter=1}
\def\aref#1;#2;#3;#4{\pp #1, {\it #2}, {\bf #3}, #4}
\def\abook#1;#2;#3{\pp #1, {\it #2}, #3}
\def\arep#1;#2;#3{\pp #1, #2, #3}

\def\Ob{\Omega_{\rm\scriptscriptstyle B}}
\def\Cth{C_{\rm th}}

\def\nhat{{\bf n}}
\def\cd{\!\cdot\!}

\def\half{{\textstyle{1\over2}}}
\def\gtwid{\mathrel{\raise.3ex\hbox{$>$\kern-.75em\lower1ex\hbox{$\sim$}}}}
\def\ltwid{\mathrel{\raise.3ex\hbox{$<$\kern-.75em\lower1ex\hbox{$\sim$}}}}

\def\Eq{Eq.$\,$}

\singlespace
\rightline{astro-ph/9402038}
\rightline{CfPA--94--TH--10}
\rightline{January 1994}

\vskip 3pt plus 0.3fill
\centerline{\titlefont Reconstructing the CMB power spectrum}

\vskip 3pt plus 0.2fill
\centerline{{\namefont Martin White}}

\vskip 3pt plus 0.1fill
\centerline{Center for Particle Astrophysics}
\centerline{University of California}
\centerline{Berkeley, CA 94720}
\centerline{USA}







{\narrower
\baselineskip=15pt
\noindent {\bf Abstract}:
We discuss a method for a model independent reconstruction of the
CMB temperature fluctuation power spectrum on small and intermediate
scales that is geared to individual experiments.
The importance of off-diagonal correlations for determining the shape
of the power spectrum is emphasized and some examples of a reconstruction
method are given.
By using this method to ``map'' the power spectrum on the scales to
which they are sensitive several experiments could be combined to
map out the full power spectrum, and test consistency of observed
features.
For example we find that the GUM scan of the 3rd flight of the MAX
experiment prefers a positive slope to the power spectrum near
$\ell\simeq160$, which provides weak evidence for the presence of
a Doppler peak.}

\vskip 4pt


\vskip 1in

\centerline{Submitted to {\it Astron. \& Astrophys.}}

\vfill\eject

\baselineskip=16pt

\vskip0.15in
\vskip\parskip
\noindent 1. {\bf Introduction}
\vskip0.1in

The study of anisotropies in the Cosmic Microwave Background (CMB)
radiation has now reached the level where experiments on a wide range
of scales are reporting detections or significant upper limits.
It seems appropriate therefore to consider ways of analyzing
the data, with a view to reconstructing the power spectrum of
temperature fluctuations.
One possibility is to give a model (e.g.~CDM) of the radiation power
spectrum in terms of a few parameters and then use many experiments to
constrain the model parameters.  In this letter we will focus on the
complimentary, model independent, approach in which a parameterization of
the power spectrum is developed on a per-experiment basis and individual
experiments are used to constrain that part of the power spectrum to which
they are most sensitive.
Since on large scales the inversion of the sky-map into a power spectrum is
a well studied problem we will focus here on individual experiments on
smaller angular scales.
Given that experiments now cover the full range of the power spectrum from
degree scales up (see e.g.~Bond~1993, White, Scott \& Silk~1994), this in
principle allows a reconstruction of the power spectrum.

\vskip0.15in
\vskip\parskip
\noindent 2. {\bf Power spectrum and window functions}
\vskip0.1in

In this letter we will assume that the CMB temerature fluctuations are
gaussian distributed so that all of the information resides in the
correlation matrix or power spectrum.  It is conventional to expand the
temperature fluctuations in spherical harmonics
$T=\sum_{\ell m}a_{\ell m}Y_{\ell m}$
and denote the power per mode by $C_\ell$:
$$
\left\langle a^{*}_{\ell m} a_{\ell' m'}\right\rangle_{\rm ens} \equiv
C_\ell\,\delta_{\ell'\ell}\delta_{m'm} \eqno(1)
$$
where the angled brackets represent an average over the ensemble of
temperature fluctuations.
The Harrison-Zel'dovich power spectrum corresponds to constant
$\ell(\ell+1)C_\ell$ or $C_\ell^{-1}\propto\ell(\ell+1)$.  In general
one expects the $C_\ell$ to be a power-law for small $\ell$ with
more structure at higher $\ell$ (smaller angular scales).
The spectrum for CDM for example, viewed as $\ell(\ell+1)C_\ell$ vs
$\ln\ell$, is a flat line until $\ell\sim100$ when it rises into two
``Doppler'' peaks before falling off exponentially at large $\ell\sim1000$
(for a discussion of power spectra in various models see
Bond \& Efstathiou~1987, Holtzman~1989, Vittorio \& Silk~1992,
Sugiyama \& Gouda~1992, Dodelson \& Jubas~1993, Stompor~1993,
Crittenden et al.~1993, Bond et al.~1993).

When fitting model parameters to data one is interesting in computing
the correlation matrix of the temperature fluctuations and from this a
likelihood function (see e.g.~Readhead \& Lawrence~1992).
The correlation matrix is a sum of two parts: the experimental error
matrix (usually assumed diagonal: $\sigma_i\delta_{ij}$) and a
``theoretical'' correlation matrix
$$\Cth(\nhat_i,\nhat_j)=
  \bigl\langle T(\nhat_i) T(\nhat_j) \bigr \rangle_{\rm ens}
 = {1\over 4\pi}\sum_{\ell=1}^\infty
   (2\ell+1)\,C_\ell\,W_\ell(\nhat_i,\nhat_j)\,, \eqno(2)$$
Here $T(\nhat)$ is the {\it measured} temperature assigned to direction
$\nhat$ and $W_\ell$ is the window function for the experiment
(e.g.~see Srednicki \& White~1994 for a discussion of window functions and
Bond~1993, White, Scott \& Silk~1994 for plots of window functions for
existing experiments).
The method proposed below does not rely on being able to construct
$W_\ell(\nhat_i,\nhat_j)$ and so is more general than just \Eq(2).
However for simplicity we will phrase our discussion as if
$W_\ell(\nhat_i,\nhat_j)$ were known, as it is for most small-scale
experiments.
For a given $\Cth$ the likelihood function is expressed in terms of
$C_{ij}=\Cth(\nhat_i,\nhat_j)+\sigma_i\delta_{ij}$ as
$$ {\cal L} \propto { 1\over\sqrt{\det C} }
  \exp\left[ -\frac1/2 T_i C_{ij}^{-1} T_j \right]  \eqno(3) $$
and provides a convenient way of using all the available experimental
information to constrain the model (in this case $\Cth$).

Usually when people discuss ``the'' window function for a particular experiment
they are referring to $W_\ell(\nhat,\nhat)$ (the window function at zero-lag)
appropriate for computing the diagonal entries of $\Cth$.
This window function defines a ``bandpass'' outside of which the experiment
has no sensitivity: the small $\ell$ cutoff is set by the amplitude of the
chop and the high $\ell$ cutoff is controlled by the beam width
(see e.g.~Bond~1993).
Integrating the power through the bandpass gives $\Delta T_{\rm rms}$,
a fact which has been exploited by several authors
(see e.g.~Bond~1993, White, Scott \& Silk~1994)
to estimate the amplitude of the power spectrum measured by the many
experiments on degree scales.
It has long been realized that separating the scales of the beam chop and
width increases the area under the window function or the sensitivity of
the experiment.

Such an increase in the width of $W_\ell$ doesn't necessarily mean less
resolving power for $C_\ell$ however.
Off-diagonal entries of $\Cth$ (i.e.~$\nhat_i\ne\nhat_j$) potentially contain
more information than just the integrated power.
They provide a lever arm which gives sensitivity to the shape of the power
spectrum, as is reflected in the shape of $W_\ell(\nhat_i,\nhat_j)$.
In figure~1 we show $W_\ell(\theta=0^{\circ})$ and $W_\ell(\theta=2^{\circ})$
($\cos\theta=\nhat_i\cd\nhat_j$, with points separated parallel to the chop
direction) for a hypothetical experiment with 2-beam, square-wave chop of
amplitude $2^\circ$ and gaussian beam of FWHM $1^\circ$.
We choose $\theta=2^\circ$ since $W_\ell(2^\circ)$ is approximately the
derivative of $W_\ell(0^\circ)$.
This derivative or``dipole-like'' form provides maximum sensitivity to
the slope of the power spectrum through the bandpass (to which $\Cth(0)$ is
mostly insensitive).

The diagonal elements of $\Cth$ coming from $W_\ell(\nhat,\nhat)$ allow an
estimate of the power through the bandpass.
{}From the above we see that without losing sensitivity to the amplitude,
the off-diagonal elements provide (model independent) information on the
{\it shape} of the power spectrum over the same range of $\ell$.

\vskip0.15in
\vskip\parskip
\noindent 3. {\bf Parameterizing the power spectrum}
\vskip0.1in

We wish to parameterize the $C_\ell$ in a simple way that is tailored to
the experiment under consideration.  Start by defining
$D_\ell\equiv\ell(\ell+1)C_\ell$ and $x\equiv\ln\ell$ and
extend $D_\ell$ to a smooth function $D(x)$ in the obvious manner.
Since on general physical grounds we expect the power spectrum to be
smooth (the neighbouring $C_\ell$ are highly ``correlated'') we can
expand $D(x)$ in a Taylor series about the peak of the window function,
$x_0$, as
$$
  D(x) = D(x_0)\left[ 1 + m (x-x_0) + \half m' (x-x_0)^2 + \cdots\right]
  \eqno(4)
$$
where $m$ and $m'$ are related to the derivatives of the power spectrum
at $x_0$.  Due to the finite beam width and chopping, any experiment is
sensitive to the power spectrum only over a limited (and usually small) range
of $x$, defined by the window function at zero-lag.
So for the purposes of each experiment, \Eq(4) should be a good expansion.
In what follows we will take as given that our expansion of $D(x)$ is not to
be extrapolated outside of the range of sensitivity of the experiment.

One problem with this expansion is that it does not enforce the physical
condition that $D(x)\ge0$ for all $x$.  Thus the allowed region of $m$,
$m'$... is constrained in a non-trivial (but calculable) way.
In principle one could get around this by expanding not $D(x)$ but $\ln D(x)$
in a power series, but this has the technical disadvantage that the expansion
is no longer linear in $D(x)$.
An expansion in which  $\Cth\propto\Cth^{(0)}(1+m\Cth^{(1)}+\cdots)$
is more computationally efficient when performing fits to the data
because $\Cth^{(0)}$, $\Cth^{(1)}$... only have to be computed once.
This has motivated us to stick with \Eq(4).
If a Monte-Carlo analysis of the data is used to constrain the parameters,
the advantage of linearity could be overlooked in favor of direct imposition
of the constraint $D\ge0$.
Eventually the data should constrain the parameters sufficiently that this
question will not be an issue.

\vskip0.15in
\vskip\parskip
\noindent 4. {\bf Two examples of the method}
\vskip0.1in

As an illustrative example of these ideas we have taken \Eq(4) and truncated
at linear order (which is all the current sensitivities merit).
Converting to the conventional notation we assume
$$ \ell(\ell+1)C_\ell = \ell_0(\ell_0+1)C_{\ell_0} \left[ 1 +
   m\ln{\ell\over\ell_0} \right]    \eqno(5) $$
where $\ell_0$ is the peak of $W_\ell(0^\circ)$.
We computed $\Cth(0^\circ,m)$ and $\Cth(2^\circ,m)$ using \Eq(5) and the
window functions of Figure~2.
We find $\Cth(0^\circ)$ (the integral through the bandpass) is mostly
insensitive to $m$ while $\Cth(2^\circ)$ has good discriminating power between
different slopes.
We have, normalizing at $m=0$,
$$ \eqalign{
\Cth(0^\circ;m)/\Cth(0^\circ;m=0) &= 1 - 0.2m\cr
\Cth(2^\circ;m)/\Cth(2^\circ;m=0) &= 1 - 1.7m\cr}  .
\eqno(6)
$$
Obviously the more off-diagonal elements one has with good signal-to-noise,
the more information one has about the shape of the power spectrum through
the bandpass, and the more one can constrain higher orders in the expansion
(the ``shape'' of the power spectrum).
In the above $\Cth(2^\circ)\simeq0.2\Cth(0^\circ)$ at $m=0$ which sets the
scale of sensitivity needed to extract the full $m$-dependence from the
signal.  If noise is a problem one could increase $\Cth(\theta)$ by measuring
correlations at other separations, but these would have reduced sensitivity
to the slope of the power spectrum.

Clearly an experiment which observes only at widely separated points will
have no non-vanishing off-diagonal correlations and sensitivity only to the
integral of the power spectrum though the bandpass.
Alternatively an experiment which samples too closely will waste time
recording ``redundant'' information and not reduce the experimental
uncertainty on the data points sufficiently to constrain the higher moments
of the power spectrum.
The question thus arises as to what observing strategy maximizes the
constraints on the power spectrum, analyzed in this manner.
Such a question cannot be answered without also considering other experimental
limitations, which will clearly vary from one experiment to another.
The question can be analyzed by the individual groups in preparing their
observing strategy.  In general one gains information on the power spectrum
by working with a strategy which (as much as possible) provides strong
off-diagonal signals.

As an application of this method to a `realistic' scenario we have computed
the likelihood function in $(C_{\ell_0},m)$ space, assuming \Eq(5), for the
GUM (Meinhold et al.~1993) scan of the 3rd flight of the MAX experiment.
The data consists of 165 points in a two-dimensional ``bow-tie'' pattern,
with varying spacing and a strongly anisotropic correlation matrix.  The
fitting procedure and data used are as discussed in Srednicki et al.~(1994).
For the parameters appropriate to MAX3 $W_\ell(\nhat,\nhat)$ peaks at
$\ln\ell_0\simeq5\ (\ell_0\simeq160)$ with a FWHM of $\ln\ell\simeq1.4$.
Due to the large number of points and the wide range of correlations
available we already find a weakly significant constraint.
[The MuPeg data set (Gundersen et al.~1993) from the same flight consisted
of relatively few, equally (``widely'') spaced data points at constant
elevation, and we find no significant constraint on $m$.]
This analysis provides {\it weak} evidence that GUM prefers a positive slope
over its range of sensitivity.
The contours of the likelihood function are shown in figure~2 (we have
restricted $m\in(-1,1)$ to enforce $C_\ell\ge0$ over the range of $\ell$
to which MAX is sensitive).
For comparison, if one weights by $(2\ell+1)^2W^2_\ell$ the $C_\ell$ for CDM
(see \Eq(2)) with $h=\half$ and $\Ob=0.01$-$0.10$, the best fits to the form
of \Eq(5) are $m=0.5$-$0.6$.
The low values of the slope come from the fact that the MAX window function
is centered near the maximum of the first ``Doppler'' peak for the models
considered here.

The analysis presented above neglects the important issue of foreground
contamination, and the data are clearly not at the level where $m$ is
severely restricted.
However it serves to show that the method works in principle.
As sensitivities increase, a more detailed analysis along these lines is sure
to provide information on the shape of the power spectrum.  A fit including
a quadratic component, for example, would allow one to test whether the power
spectrum exhibits a peak in the range of $\ell$ probed (at present the
likelihood function for MAX is very broad in $m'$ and no conclusion can
be drawn).

\vskip0.15in
\vskip\parskip
\noindent 5. {\bf Conclusions}
\vskip0.1in

In the future, with reduced experimental errors and more coverage, the
method presented in this letter could be used to map the power spectrum,
over the range of sensitivity of individual experiments, in a way
which provides more and more shape information as higher off-diagonal
correlations are observed above the noise.
Pasting together constraints of this form from many experiments should
allow one to reconstruct the full power spectrum in a model independent
manner.

\vskip0.15in
\vskip\parskip
\noindent {\bf Acknowledgements}
\vskip0.1in

I would like to thank Emory Bunn, Joanne Cohn, Douglas Scott, Joe Silk
and Mark Srednicki for many useful conversations and comments on this and
related work.
This research was supported by the NSF and the TNRLC.


\vskip0.4in
\noindent {\bf References}
\vskip0.1in
\frenchspacing
\parindent=0truept

\abook Bond, J. R., 1993;{\rm in} Proceedings of the IUCAA Dedication
Ceremonies;ed. T. Padmanabhan, John Wiley \& Sons, New York, in press

\aref Bond, J. R. \& Efstathiou, G., 1987;MNRAS;226;655

\arep Bond, J. R., Crittenden, R., Davis, R. L., Efstathiou, G.,
Steinhardt, P. J.,~1993;preprint;Penn

\aref Crittenden, R., Davis, R. L.~\& Steinhardt, P. J., 1993;Ap. J.;417;L13

\aref Dodelson, S. \& Jubas, J. M., 1993;Phys. Rev. Lett.;70;2224

\aref Gundersen, J. O., et al., 1993;ApJ;413;L1

\aref Holtzman, J. A., 1989; ApJ (Supp);71;1

\aref Meinhold, P. R., et al., 1993;ApJ;409;L1

\aref Readhead, A. C. S. \& Lawrence, C. R., 1992;Ann Rev Astron \&
Astrophys;30;653

\abook Srednicki, M., White, M., Scott, D.~\& Bunn, E.,
1994;Phys Rev Lett;to appear

\abook Srednicki, M.~\& White, M., 1994;CfPA preprint;in preparation

\abook Stompor, R., 1993;Astron Astrophys;in press

\aref Sugiyama, N. \& Gouda, N., 1992;Prog. Theor. Phys.;88;803

\aref Vittorio, N. \& Silk, J., 1992;ApJ;385;L9

\abook White, M., Scott, D.~\& Silk, J., 1994;Ann Rev Astron \& Astrophys;
in press

\nonfrenchspacing

\vskip0.15in
\vskip\parskip
\noindent {\bf Figure Caption}
\vskip0.1in
\parindent=0pt

Fig.~1.\ The window functions $W_\ell(\theta=0^\circ)$ (solid) and
$W_\ell(\theta=2^\circ)$ (dashed) for a hypothetical experiment with
2-beam, square wave chop of amplitude $2^\circ$ and a gaussian beam of
FWHM $1^\circ$.

Fig.~2.\ The contours of the likelihood function in $(C_{\ell_0},m)$ for
the MAX-GUM data.  The contours are in units of $\half\sigma$.

\end